\documentclass[%
 reprint,
 amsmath,amssymb,
 aps,
]{revtex4-2}

\usepackage{graphicx}
\usepackage{dcolumn}
\usepackage{bm}
\usepackage{hyperref}

\usepackage{xcolor}
\definecolor{myBlue}{rgb}{0,0.4314,0.7333}
\definecolor{myGreen1}{rgb}{0,0.69,0.3137}
\definecolor{myGreen2}{rgb}{0,0.4667,0}
\definecolor{myPurple1}{rgb}{0.3725,0.4588,0.9882}
\definecolor{myPurple2}{rgb}{0.4196,0,0.5255}

\begin{document}

\preprint{APS/123-QED}

\title{Depolarization composition of the back-scattered circularly polarized light}

\author{Ivan Lopushenko}
\affiliation{Opto-Electronics and Measurement Techniques, ITEE, University of Oulu, Oulu, 90014, Finland}
\author{Alexander Bykov}
\affiliation{Opto-Electronics and Measurement Techniques, ITEE, University of Oulu, Oulu, 90014, Finland}
\author{Igor Meglinski}
\email{Correspondence: i.meglinski@aston.ac.uk}
\affiliation{%
 College of Engineering and Physical Sciences, Aston University, Birmingham, B4 7ET, UK 
}

\date{\today}

\begin{abstract}
We consider the origin of unpolarised light resulting from the backscattering of circularly polarized light by random turbid tissue-like disperse medium. We reveal the dynamics of the backscattered fraction of unpolarized light, disclosing its meticulous decomposition into two rigorously polarized components characterized by opposing helicities, with fully defined polarization states. Concurrently, their superposition, driven by multiple scattering within the medium, leads to the appearance of a fraction of linear polarization. We emphasize that in-depth binding of circular polarization memory of light with the helicity flips occurs within the scattering medium, meaning the conservation of spin angular momentum. We anticipate that the results obtained hold significant implications for future studies, particularly in the field of tissue polarimetry and light vortexes.
\end{abstract}
\maketitle

In conjunction with the wavelength and coherence the polarization is a fundamental property of light~\cite{Goldstein2003} attracting great attention in numerous practical biomedical and clinical applications~\cite{JBO2016,BOE2021,Booth}. The circularly polarized light carries an intrinsic angular momentum due to its polarization helicity~\cite{Cameron}. Recent studies have shown that phase retardation of circularly polarized light, back-scattered by biological tissue, can be used effectively for quantitative evaluation of cervical intraepithelial neoplasia~\cite{Kinunen2015}, presence of senile Alzheimer's plaques~\cite{Borovkova2020,Borovkova2022} and characterization of biological tissues with optical anisotropy~\cite{Ushenko2019,Sieryi2022}. This polarimetry approach is based on the directional awareness of circularly polarized light, when by known stage of polarization of incident light the helicity of scattered light can be used to determine if it has been forward or back scattered~\cite{Vitkin23}. This peculiar property of circularly polarized light is also known as a polarization memory~\cite{MacKintosh,Alfano2005}. Directional awareness of circularly polarized light is a manifestation of anisotropy of scattering~\cite{Macdonald2015}.
Linear polarization possesses no such sense of the directional awareness. Scattered multiple number of times in turbid tissue-like disordered medium linearly or circularly polarized light is depolarized, and the depolarization degree depends strongly on the size and shape of scattering particles~\cite{Bicout,Hielscher1997}, as well as on the number of scattering events~\cite{Lenke}. 
Quantitatively the residual state of polarization is defined as the ratio of polarized intensity to the total intensity of back-scattered light, known as a degree of polarization ($DoP$). $DoP$ is defined utilizing Stokes vector parameters~\cite{Goldstein2003}:
\begin{equation}
    \label{eqn:DOP}
    DoP = \dfrac{\sqrt{S_1^2+S_2^2+S_3^2}}{S_0}. 
\end{equation}
Here, $S_1$ is the difference in intensity between horizontally and vertically polarized light $(S_1=I_{\parallel}-I_{\perp})$, $S_2$ is the difference in intensity between linearly polarized light observed at $45^\circ$ and $-45^\circ$ $(S_2=I_{+45^\circ}-I_{-45^\circ})$, $S_3$ is the difference in intensity between left- and right- circularly polarized light ($S_3=I_{L}-I_{R}$), and $S_0$ is the total intensity of light ($S_0 = I_{\parallel}+I_{\perp}=I_{+45^\circ}+I_{-45^\circ}=I_{L}+I_{R}$). 

The circularly polarized light back-scattered from a turbid tissue-like disordered medium is represented as a sum of two Stokes vectors defining unpolarized and totally polarized light~\cite{Goldstein2003}: 
\begin{equation}
\label{eq:StokesdecoupleIntoUnpolAndPol}
    \begin{array}{c}
\left(\begin{array}{c}
    S_0\\ 
    S_1\\ 
    S_2\\ 
    S_3\\
    \end{array}\right) = (1-DoP)\left(\begin{array}{c}
    S_0\\ 
    0\\ 
    0\\ 
    0\\
    \end{array}\right) + \left(\begin{array}{c}
    DoP \cdot S_0\\ 
    S_1\\ 
    S_2\\ 
    S_3\\
    \end{array}\right). \\ 0\leq DoP\leq1.\end{array}
\end{equation}
In addition to the overall $DoP$, there are specific measures of the degree for different types of polarization: 
\begin{equation}
    \label{eqn:DOLP}
DoLP = \dfrac{\sqrt{S_1^2+S_2^2}}{S_0},
\end{equation}
and
\begin{equation}
    \label{eqn:DOCP1}
DoCP = \dfrac{\sqrt{S_3^2}}{S_0}.
\end{equation}
Here, $DoLP$ and $DoCP$ are, respectively, the degree of linear polarization and the degree of circular polarization ($DoP^2 = DoLP^2 + DoCP^2$). In practice $DoCP$ is often presented as~\cite{Born2019}:
\begin{equation}
    \label{eqn:DOCP2}
DoCP = \dfrac{S_3}{S_0} = \dfrac{{I}_{L}-{I}_{R}}{{I}_{L}+{I}_{R}},
\end{equation}
known also as the Circular Intensity Differential Scattering (CIDS)~\cite{Bustamante1980}, showing the degree of differential scattering between measured intensities of left- and right- circular polarization, and provides a quantitative measure of  polarization-preserving properties of scattering medium. 

Partially polarized light is decomposed in two completely polarized fractions of light with opposite helicity~\cite{Goldstein2003}:
\begin{widetext}
\begin{eqnarray}   \label{eq:StokesdecoupleIntoTwoOppositePol}
\left(\begin{array}{c}
    S_0\\ 
    S_1\\ 
    S_2\\ 
    S_3\\
    \end{array}\right) &=& \frac{(1+DoP)}{2DoP}\left(\begin{array}{c}
    DoP \cdot S_0\\ 
    S_1\\ 
    S_2\\ 
    S_3\\
    \end{array}\right)+ 
    \frac{(1-DoP)}{2DoP}\left(\begin{array}{c}
    DoP \cdot S_0\\ 
    -S_1\\ 
    -S_2\\ 
    -S_3\\
    \end{array}\right). 
\end{eqnarray}
\end{widetext}
From the practical point of view the expression (\ref{eq:StokesdecoupleIntoUnpolAndPol}) is represented in terms of Stokes parameters normalized to the intensity of the fully polarized component, i.e.: 
\begin{equation}
\label{eq:StokesNormalized}
Q = \dfrac{S_1}{DoP \cdot S_0}, \: U = \dfrac{S_2}{DoP \cdot S_0}, \: V = \dfrac{S_3}{DoP \cdot S_0}.
\end{equation} 
This allows to assess quantitatively the parameters of Stokes vector observed by conventional polarimeter~\cite{Thorlabs}. 
Thus, for the light depolarized due to propagation through a turbid tissue-like scattering medium (\ref{eq:StokesdecoupleIntoUnpolAndPol}) takes the form
\begin{equation}
\label{eq:StokesdecoupleIntoUnpolAndPolNorm1}
    \begin{array}{c}
\left(\begin{array}{c}
    S_0\\ 
    S_1\\ 
    S_2\\ 
    S_3\\
    \end{array}\right)
 = (1-DoP)\, S_0\left(\begin{array}{c}
    1\\ 
    0\\ 
    0\\ 
    0\\
    \end{array}\right) + DoP \, S_0 \left(\begin{array}{c}
    1\\ 
    Q\\ 
    U\\ 
    V\\
\end{array}\right)\end{array}
\end{equation}
In turbid scattering medium, such as biological tissue, $DoP$ drop can be significant, leading to a significant decrease of the signal-to-noise ratio and a loss of sensitivity and accuracy in optical measurements~\cite{Booth}. 

By analogy to  (\ref{eq:StokesdecoupleIntoTwoOppositePol}) partially depolarized ($0 < DoCP < 1$) back-scattered circularly polarized light is presented in terms of specific polarization measure (\ref{eqn:DOCP1}) as a sum of totally polarized and completely unpolarized light that according to (\ref{eq:StokesdecoupleIntoUnpolAndPolNorm1}) can be also decomposed into two opposite polarization states:
\begin{widetext}
\begin{eqnarray}
\left(\begin{array}{c}
    S_0\\ 
    0\\ 
    0\\ 
    S_3\\
   \end{array}\right)&=&DoCP \cdot S_0
    \left(\begin{array}{c}
    1\\ 
    0\\ 
    0\\ 
    1\\
    \end{array}\right)+ 
    \left(1-DoCP\right)\left(\dfrac{S_0}{2}\left(\begin{array}{c}
    1\\ 
    0\\ 
    0\\ 
    -1\\
    \end{array}\right) 
    + \dfrac{S_0}{2}\left(\begin{array}{c}
    1\\ 
    0\\ 
    0\\ 
    1\\
\end{array}\right)\right). 
\end{eqnarray}
Thus, back-scattered from the medium circularly polarized light can be represented as a decomposition of two totally polarized left circularly polarized ($LCP$) and right circularly polarized ($RCP$) components:
\begin{eqnarray}
\label{eq:DOCP1}
\left(\begin{array}{c}
    S_0\\ 
    0\\ 
    0\\ 
    S_3\\
    \end{array}\right) &=& 
    \dfrac{(1-DoCP) \cdot S_0}{2}\left(\begin{array}{c}
    1\\ 
    0\\ 
    0\\ 
    -1\\
    \end{array}\right)+ 
    \dfrac{(1+DoCP) \cdot S_0}{2}
    \left(\begin{array}{c}
    {1}\\ 
    0\\ 
    0\\ 
    1\\
    \end{array}\right).
\end{eqnarray} 
In a similar way the depolarized linearly polarized light can be also presented as a decomposition of linear horizontal $I_{\parallel}$ and vertical $I_{\perp}$ states of polarization:
\begin{eqnarray}
\label{eq:DOLPdecomposition1}
\left(\begin{array}{c}
    S_0\\ 
    S_1\\ 
    S_2\\ 
    0\\
    \end{array}\right) &=& 
    \frac{(1+DoLP) \cdot S_0}{2}\left(\begin{array}{c}
    1\\ 
    Q\\ 
    U\\ 
    0\\
    \end{array}\right)+ 
    \frac{(1-DoLP) \cdot S_0}{2}\left(\begin{array}{c}
    1\\ 
    -Q\\ 
    -U\\ 
    0\\
    \end{array}\right),
\end{eqnarray}
that reduces accordingly to:
\begin{eqnarray}
\label{eq:DRdecomposition}
\left(\begin{array}{c}
    S_0\\ 
    S_1\\ 
    0\\ 
    0\\
    \end{array}\right) 
    &=& \frac{(1+DR) \cdot S_0}{2}\left(\begin{array}{c}
    1\\ 
    1\\ 
    0\\ 
    0\\
    \end{array}\right)+ 
    \frac{(1-DR) \cdot S_0}{2}\left(\begin{array}{c}
    1\\ 
    -1\\ 
    0\\ 
    0\\
    \end{array}\right), 
\end{eqnarray}
\end{widetext}
if $S_2=0$. Here, 
\begin{eqnarray}
\label{eq:DR}
 DR = \sqrt{S_1^2}/S_0 =  \dfrac{{I}_{\parallel}-{I}_{\perp}}{{I}_{\parallel}+{I}_{\perp}} 
\end{eqnarray}
is the depolarization ratio~\cite{Lenke} for linearly polarized light, defined as the ratio of the intensity of polarized light to the total intensity.

\begin{figure}[!ht]
\includegraphics[width=.9\linewidth]{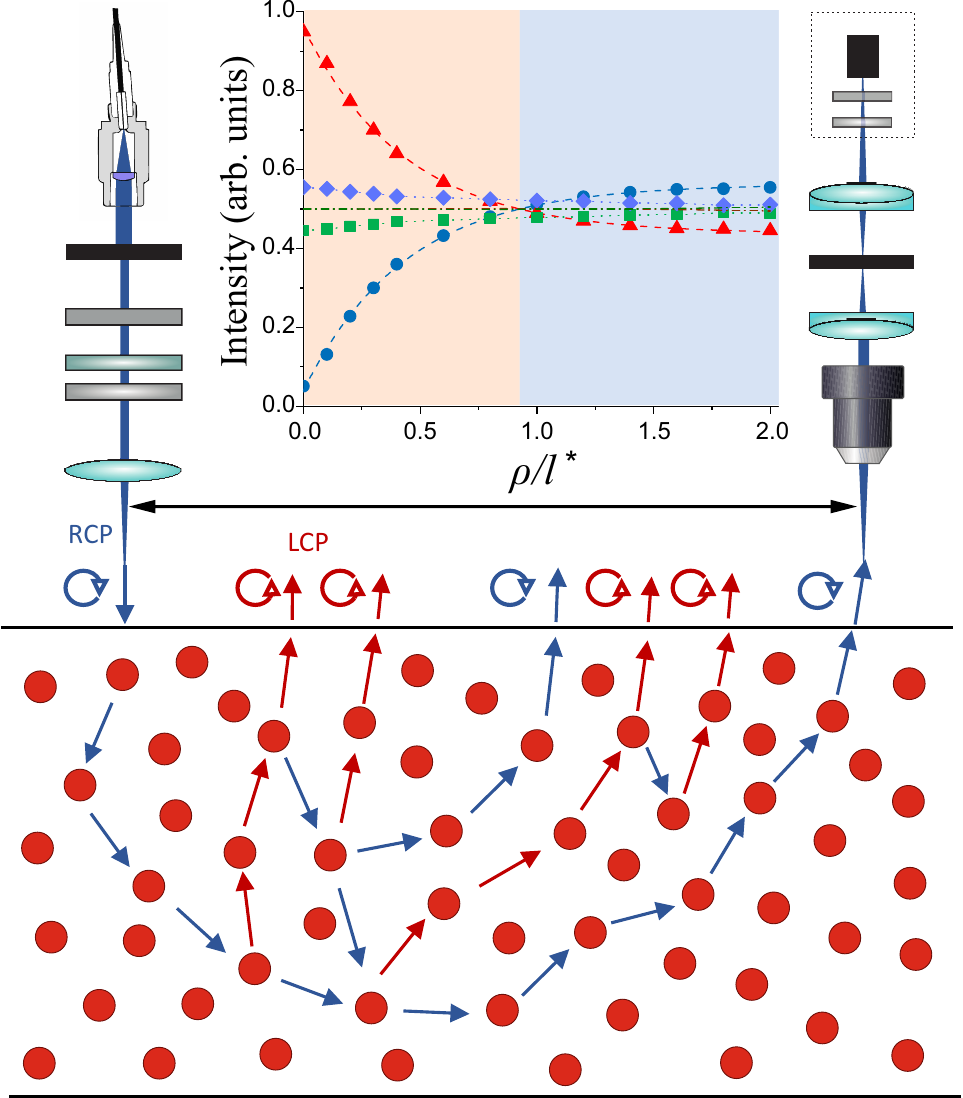}
\caption{\label{fig:vertical_intensity} Schematic presentation of the experimental setup, along with selected potential photon trajectories within the medium, wherein each backscattering event induces a helicity flip, visually depicted by the color of the directional arrow. The inset shows $LCP$ ({\color{red}$\blacktriangle$}) and $RCP$ ({\color{myBlue}$\bullet$}) components of the back-scattered $RCP$ light, and the pairs of orthogonal intensities $I_{\parallel}$ ({\color{myGreen1}$\blacksquare$}), $I_{\perp}$ ({\color{myPurple1}$\diamond$}) and $I_{+45^\circ}$ (black line), $I_{-45^\circ}$ (dark green line) depending on the scaled source-detector separation $\rho/l^*$.}
\end{figure}

The experimental approach commonly employed for direct measurements of the intensity of the back-scattered polarized light, as well as $DoP, DoLP, DoCP, DR$ quantities~\cite{Macdonald2011,Kinunen2015,Vitkin23,Ivanov2020,Borovkova2020,Borovkova2022}, is depicted schematically in Figure~\ref{fig:vertical_intensity}. The transformation of low-coherent laser light into $RCP$, achieved using a set of lenses, half-wave, and quarter-wave plates, facilitates its focused delivery onto the turbid tissue-like scattering medium or tissue sample. The photons entered the medium undergo a sequence of scattering events before they are detected. Back-scattered light is collected by an objective positioned at a particular distance $\rho$ from the point of incidence and subsequently passed through an analyzer to measure its polarization state. The trajectories of photons within the medium, schematically shown in Figure \ref{fig:vertical_intensity}, are determined by a random walk characterized by a mean free path $l$, ultimately culminating in their exit point, where the measured state of polarization is determined.
 
In this study, we consider the placement of the light source and detector to be perpendicular to the surface of the semi-infinite homogeneous turbid tissue-like scattering medium. The following parameters of the medium are used both in experiment and theoretical modeling: scattering coefficient $\mu_s=4\,\,mm^{-1}$ ($\mu_s = 1/l$), absorption coefficient $\mu_a=0.05\,\,mm^{-1}$, anisotropy scattering factor $g = 0.9$, refractive index $n=1.46$ at the wavelength of incident light $\lambda=640\,\,nm$ ($l \gg \lambda$); $g \equiv \langle cos\,\theta \rangle$, where $\theta$ is the scattering angle and the average $\langle ... \rangle$ is taken over the form factor of the medium scattering particle. In our experimental investigations, phantoms possessing the aforementioned optical properties are meticulously designed and crafted in accordance with the established manufacturing protocols~\cite{Sieryi2020,Jessica2019}.

Figure \ref{fig:vertical_intensity} also illustrates the interplay between the oppositely polarized components of the detected light.
The $\rho$ is scaled to the transport mean free path $(l^*=1/(1-g)$, the average distance that light propagate before it direction of propagation is totally randomized~\cite{Ishimaru:book}. 

As evident at the inset of Figure 1, for short source-detector separations ($\rho/l^*<1$), the helicity of the incident $RCP$ light undergoes flipping as a result of backscattering. The flipped $LCP$ light is inversely related to the emerging $RCP$ component (see Fig.~\ref{fig:vertical_intensity}). The $LCP$ light is formed due to odd number of the helicity flips occurred along the consecutive scattering events within the medium between points of incidence and detection, whereas appearance of $RCP$ is based on the even number of flips~\cite{MacKintosh}. The decrease of $LCP$ with the increase source-detector separation is compensated with the proportional increase of $RCP$ light (see Fig.~\ref{fig:vertical_intensity}), clearly illustrating predictions (\ref{eq:DOCP1}). The $RCP$ stream becomes dominating over $LCP$ at larger source-detector separation ($\rho > l^*$), meaning that the conservation of angular momentum is preserved, and multiple scattering maintain the helicity of incident circularly polarized light, i.e. $RCP$. At the isosbestic point (see Fig.~\ref{fig:vertical_intensity}) the intensities of two streams of light with opposite helicities are equalized ($I_L = I_R$) and their superposition originates linear polarization. The orthogonal linearly polarized components: $I_{\parallel},I_{\perp}$ and
$I_{+45^\circ},I_{-45^\circ}$, engendered due to scattering of incident $RCP$ light by mean of a superposition of the fully polarized, flipped, $LCP$ light and the appeared with opposite helicity $RCP$ component, are also presented in Figure~\ref{fig:vertical_intensity}.
\begin{figure}[!h]
\includegraphics[width=.9\linewidth]{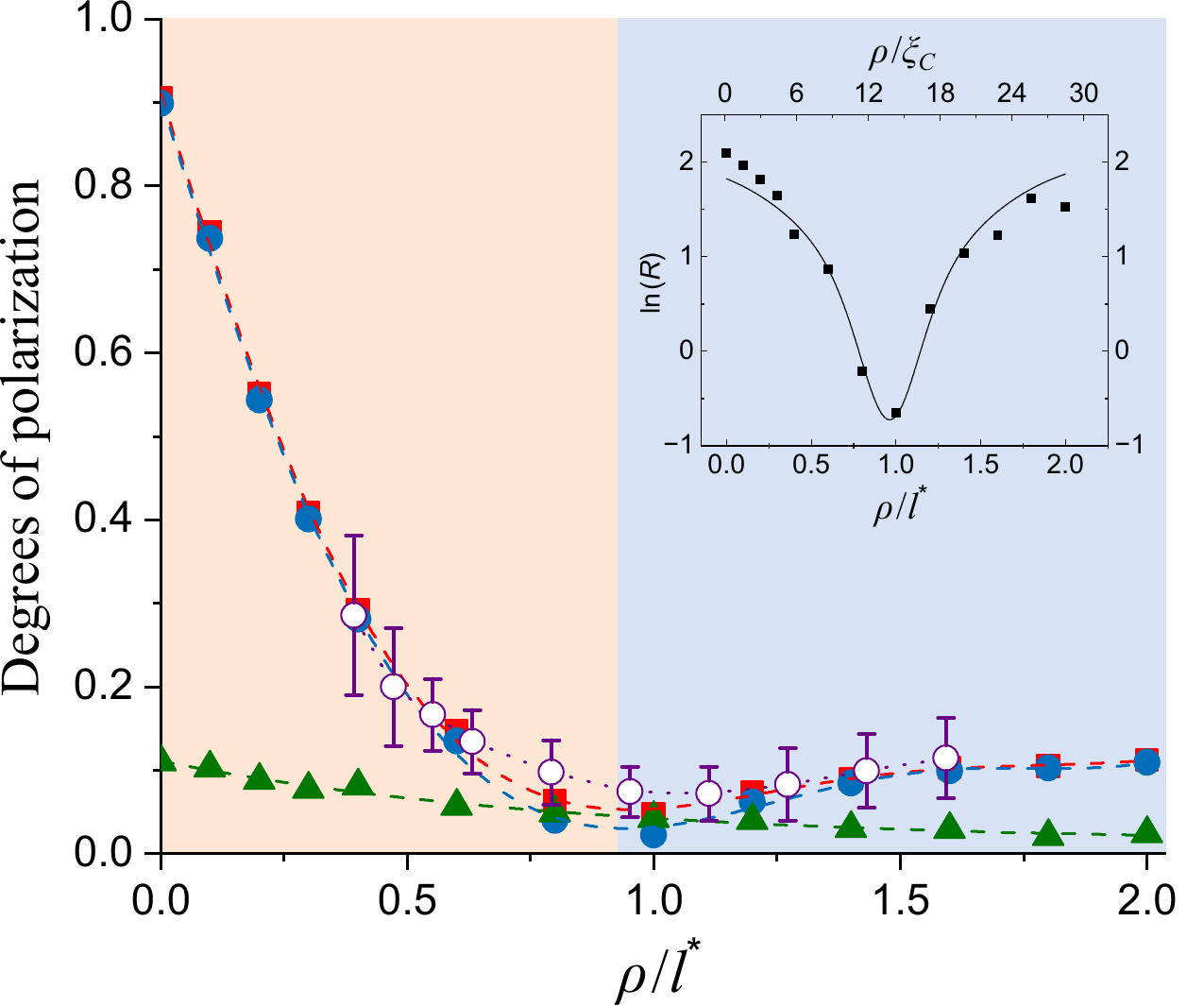}
\caption{\label{fig:vertical_DOP} 
$DoP$ ({\color{red}$\blacksquare$}), $DoCP$ ({\color{myBlue}$\bullet$}), and $DoLP$ ({\color{myGreen2}$\blacktriangle$}) plotted as function of scaled source-detector separation $\rho/l^*$. Purple line ({\color{myPurple2}$\circ$}) corresponds to the experimental data. Inset depicts the ratio of circular to the linear degree of polarization $R = DoCP/DoLP$ as a function of $\rho/l^*$ (bottom) and/or by source-detector separation scaled by characteristic depolarization length ($\rho/\xi_C$) for the circularly polarized light (top).}
\end{figure}

The polarization memory is revealed as a flip of the back-scattered circularly polarized light at the source-detector separation over the transport length ($\rho > l^*$), tailing the helicity of incident $RCP$ light (see Fig.~\ref{fig:vertical_intensity}). 

The results presented in Figure~\ref{fig:vertical_intensity} and figures below 
are attained using a Monte Carlo (MC) modeling approach grounded in the iterative procedure of the solution of the Bethe-Salpeter (BS) equation~\cite{Meglinski}. The merits of this MC methodology predicated encompass a direct correspondence with the analytic Milne solution~\cite{Kuzmin2007} and an inherently comprehensible physical elucidation of the coherent effects of multiple scattering of light through the utilization of ladder diagrams~\cite{Meglinski}. By employing the Jones vector formalism, this numerical scheme has been established as efficacious for tracking the polarization of photons within a turbid tissue-like medium and for simulating coherent back-scattering phenomena~\cite{Doronin2014a}. The validity of the approach has been justified on the fundamental level~\cite{DoicuMishchenko2019}. 

A comparison of $DoP$, $DoLP$ and $DoCP$ towards scaled source-detector separation $\rho/l^*$ are presented in Figure \ref{fig:vertical_DOP}. $DoCP$ represents the fraction of the circularly polarized light that is preserved or retained after the multiple scattering. With the increase of source-detector separation the $DoCP$ is decreased due to reduction of low scattering orders contribution to the back-scattered light. At a particular source-detector separation $(\rho/l^*)$ where flipped $I_L$ and preserved $I_R$ components of the back-scattered circularly polarised light are equalized (see Fig.~\ref{fig:vertical_DOP}), the $DoCP$ reaches a minimum value. The minimum of $DoLP$ is observed at the same point due to parities of pairs $I_{\parallel},I_{\perp}$ and $I_{+45^\circ},I_{-45^\circ}$ (see Fig.~\ref{fig:vertical_intensity}). As a result $DoP$ reaches minimal value at the same distance as well (see Fig.~\ref{fig:vertical_DOP}). The depolarization minimum represents the point at which the components of scattered circularly light with opposite helicity, $LCP$ and $RCP$, are superimposed. The depolarization minimum is coincided with the demarcation line between non-diffusive and diffusing path-lengths of scattering photons characterized by $l^*$ (see Fig.~\ref{fig:vertical_DOP}). These results are in a good agreement with the results of experimental studies performed earlier~\cite{Borovkova2022}, as shown in Figure~\ref{fig:vertical_DOP}. The inset presented in Figure \ref{fig:vertical_DOP} shows the ratio of circular to the linear degree of polarization $R = DoCP/DoLP$ as a function of $\rho/l^*$ and the source-detector separation scaled by characteristic depolarization length for circularly polarized light $\xi_C$~\cite{PhysRevE.49.1767,PhysRevE.64.026612}. This, so-called circular depolarization ratio (CDR)~\cite{Radio}, specifically quantifies the balance between circularly polarized and linearly polarized components of light within the multiple scattering. The inset results show that the portion of the backscattered light that retains its circular polarization versus the portion that becomes linearly polarized is dropped down quickly and reversely raised upon the helicity flip (see Fig.~\ref{fig:vertical_DOP}). 

\begin{figure}[!h]
\includegraphics[width=.85\linewidth]{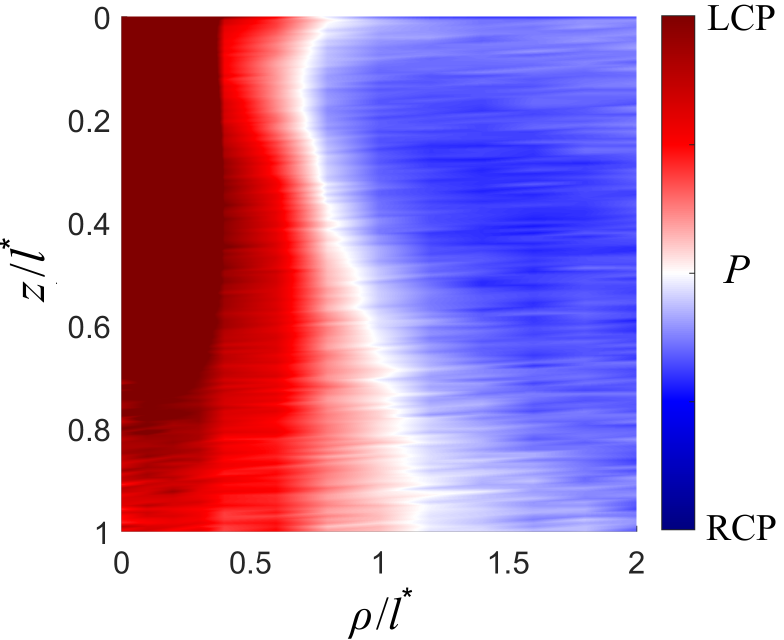}
\caption{\label{fig:vertical_memory}In-depth spatial distribution of the polarization memory within the turbid-tissue-like disordered medium in terms of PRE between flipped $LCP$ and preserved $RCP$ components of the back-scattered circularly polarized light; $z/l^*$ is the dimensionless depth penetration and $\rho/l^*$ is the scaled source-detector separation.}
\end{figure}

The extent of polarization cross talk between flipped and preserved components of the back-scattered circularly polarized light is characterized by the polarization extinction ratio (PER)~\cite{Azzam:book}: $(P = I_L/I_R)$. Figure \ref{fig:vertical_memory} shows the in-depth spatial distribution of the polarization memory, presented by analogy to the photon-measurement density function (PMDF)~\cite{Arridge}, in terms of gradient of PRE ($\nabla P_{LCP/RCP}\left(\mathbf{r}\right)=\partial P/\partial x,\partial P/\partial y,\partial P/\partial z$) at each pixel ($\mathbf{r} = (x,y,z)$) in the medium: 
\begin{equation}
P_{LCP/RCP}\left(\mathbf{r}\right)=\frac{\sum\limits_{i=1}^{N_{ph}} s_i\left(\mathbf{r}\right) I_{L|R}}{s_0 \sum\limits_{i=1}^{N_{ph}} I_{L|R}}.
\label{eq:svol}
\end{equation}
Here, $I_{L|R}$ corresponds to the detected $LCP$ and/or $LCP$ intensities, $s_i\left(\mathbf{r}\right)$ is the path length of the $i$-th photon within a pixel centered at $\mathbf{r}$, $s_0$ is the linear size of pixel, $N_{ph}$ is the total number of detected photons. 
The in-depth spatial distribution (see Fig.~\ref{fig:vertical_memory}) shows a strong localization of $LCP$ component in relation to the incident polarization state at the short ($\rho < l^*$) source-detector distances. The linear polarization, emerged as a superposition of $LCP$ and $RCP$ components, demarcates areas of their localization. The wide aperture of light source ($NA \sim 70^\circ$) and anisotropy of scattering ($g$) result in a broad range of scattering angles of photons and their pathlength distribution, leading to an asymmetry of the in-depth spatial distribution.

\begin{figure}[!h]
\includegraphics[width=.9\linewidth]{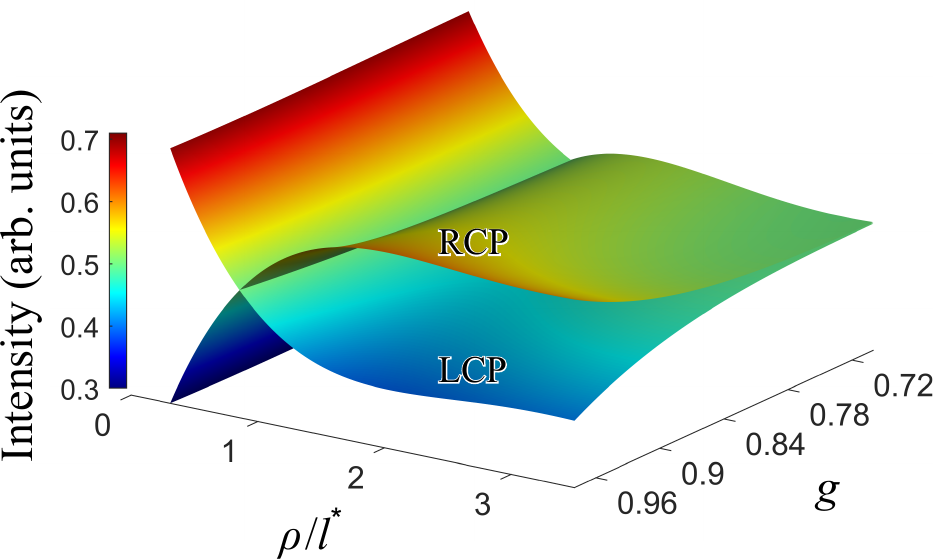}
\caption{\label{fig:coCross3D} $LCP$ and $RCP$ intensities of the backscattered light as a function of anisotropy of scattering $g$ and source-detector separation scaled to transport length $l^*$. }
\end{figure}

It should be pointed out that depolarization composition of the backscattered circularly polarized light varies depending on the properties of turbid tissue-like disperse medium, such as its scattering characteristics, the size and composition of scattering particles, and the overall optical density~\cite{Bicout,Hielscher1997,Lenke}. In this point of view, as an example, Figure \ref{fig:coCross3D} shows $LCP$ and $RCP$ components depending on the anisotropy of scattering ($g$); a deviation of the isosbestic point ($I_{L}=I_{R}$) is clearly observed. 

To sum up, we focused attention on an important quantity of interest: the depolarization of circularly polarized light back-scattered from semi-infinite turbid tissue-like disperse medium. When the circularly polarized light undergoes multiple scattering events within a turbid tissue-like disordered medium, it follows a convoluted and intricate path due to interactions with the scattering particles and structures within the medium. In biological tissues these scattering particles and structures include cellular components, extracellular matrices, organelles, and other microstructures. It is generally assumed that light multiply scattered from turbid tissue-like disordered medium randomizes the state of initial polarization. Multiple scattering events cause a loss of coherence of incident light. Therefore, phase relationships between different scattered waves become random, leading to the loss of initial polarization state. In other words, after undergoing multiple scattering, the polarization state of light becomes depolarized, exhibiting a mixture of polarization states. While multiple scattering tends to depolarize light, circular polarization exhibits a higher degree of preservation due to preferential scattering interactions. As a result, despite of multiple scattering the polarization memory of circularly polarized light is observed over a depolarization framework. We explore the evolution of polarisation memory of circularly polarized light back-scattered from turbid tissue-like disordered medium utilizing Stokes vector formalism. More specifically, we address the in-depth binding of circular polarization memory with the helicity flips occurring within the medium. We show that for normal incidence and detection of circularly polarized light the flipped helicity survival is prevailed at the short source-detector separation ($\rho < l^*$). A transition from $LCP$ to $RCP$ is revealed for longer distances ($\rho > l^*$) resultant preservation of the helicity of incident light. We show that back-scattered circularly polarized light is decomposed into two fully polarized components with opposite helicities, and their polarization states are fully defined. Thus, the depolarization composition of the backscattered circularly polarized light refers to the fully polarized components of opposite helicity, their superposition resulting in some linear polarization, with no contribution from unpolarized light. 

\begin{acknowledgments}
The authors express their gratitude to Dr. Tatiana Novikova from LPICM, Ecole polytechnique (France), for her invaluable contributions through insightful discussions and critical comments during the course of this study and the preparation of this paper. Authors acknowledge the support from ATTRACT II META-HiLight project funded by the European Union’s Horizon 2020 research and innovative programme under grant agreement No.101004462, the Academy of Finland (grant project 325097), the Leverhulme Trust and The Royal Society (Ref. no.: APX111232 APEX Awards 2021), and the Universities UK International (UUKi) and the DSTI – UK government Department for Science, Innovation and Technology (Light4Body grant project).
\end{acknowledgments}

\bibliography{references}

\end{document}